\documentclass[aps,preprint,showpacs,floatfix]{revtex4-1}
\usepackage{bm}
\usepackage{amsmath}
\usepackage{epsfig}
\usepackage{graphicx}
\usepackage{subfigure}

\def\pr{\prime}
\def\be{\begin{equation}}
\def\lan{\left\langle}
\def\ran{\right\rangle}
\def\ee{\end{equation}}
\def\barr{\begin{array}}
\def\earr{\end{array}}

\def\nn8{\\}
\def\l{\left}
\def\r{\right}
\def\dis{\displaystyle}
\def\ed{\end{document}}

\def\bF{{\mbox{\boldmath $F$}}}

\def\cn{{\cal N}}

\def\cq{{\cal Q}}

\def\car{{\cal R}}

\def\cs{{\bf s}}

\def\wh{{\widehat {H}}}

\begin{document}

\title{Embedded Gaussian Unitary Ensembles with $U(\Omega) \otimes SU(r)$
Embedding generated by Random Two-body Interactions with $SU(r)$ Symmetry}

\author{Manan Vyas$^{1,}$\footnote{ Corresponding author, phone: 509-335-4675,
Fax: 509-335-7816 \\ {\it E-mail address: manan.vyas@wsu.edu} (Manan Vyas)} and  V.K.B.
Kota$^{2}$\ }

\affiliation{$^1$Department of Physics and Astronomy, Washington State
University, Pullman, Washington 99164-2814, USA \\ 
$^2$Physical Research Laboratory, Ahmedabad 380 009, India  }

\begin{abstract}

Following the earlier studies on embedded unitary ensembles generated by random
two-body interactions [EGUE(2)] with spin $SU(2)$ and spin-isospin $SU(4)$
symmetries, developed is a general formulation, for deriving lower order 
moments of the one- and two-point correlation functions in eigenvalues, that is
valid for any EGUE(2) and BEGUE(2) ('B' stands for bosons)  with $U(\Omega)
\otimes SU(r)$ embedding and with two-body interactions preserving $SU(r)$
symmetry. Using this formulation with $r=1$, we recover the results derived by
Asaga et al [Ann. Phys. (N.Y.) {\bf 297}, 344 (2002)] for spinless boson
systems. Going further, new results are obtained for $r=2$ (this corresponds to
two species boson systems) and $r=3$ (this corresponds to spin 1 boson
systems). 

\end{abstract}


\maketitle

\section{Introduction}

A long standing question for the embedded ensembles is about their analytical
tractability. Amenability to mathematical treatment is one of the four
conditions laid down by Dyson \cite{Dy-72} for the validity of a random matrix
ensemble. Simplest of the two-body unitary ensemble is the embedded  Gaussian
unitary ensemble of  two-body interactions [EGUE(2)] for spinless fermion
systems. For $m$ fermions in $N$ sp states, the embedding is generated by the 
$SU(N)$ algebra and although this ensemble is known for many years,  only
recently \cite{Ko-05}, after the first indications  implicit in
\cite{Be-01b,PlW}, it is established that the $SU(N)$  Wigner-Racah algebra
solves EGUE($2$) and also the more general  EGUE($k$) [as well as EGOE($k)$].
These results, with $U(N)$ algebra, extended to BEGUE($k$) for spinless bosons in
$N$ sp states (see \cite{Ko-05,Asa-02}). For EGUE(2)-$\cs$ for fermions with
spin and EGUE(2)-$SU(4)$  for fermions with Wigner's spin-isospin $SU(4)$
symmetry, the embedding algebras, with $\Omega$ number of spatial degrees of
freedom for a single fermion, are $U(\Omega) \otimes SU(2)$ and $U(\Omega)
\otimes SU(4)$ respectively. It was shown in \cite{Ko-07,Ma-su4} that the
Wigner-Racah algebra of these embedding algebras will allow one to obtain
analytical results for the lower order moments of the one- and two-point
correlation functions in eigenvalues. Similarly, following the recent work
\cite{Ma-12,Ckmp-arx} on BEGOEs, it is easy to recognize that the embedding
algebras for BEGUE(2)-$F$ for two-species boson systems   with $F$-spin and
BEGUE(2)-$SU(3)$ for spin one boson systems are  $U(\Omega) \otimes  SU(2)$ and
$U(\Omega) \otimes SU(3)$ respectively. The  purpose of the present paper is to 
establish on one hand that the  Wigner-Racah algebra of these embedding algebras
solve the corresponding  embedded unitary ensembles and on the other to generalize the
formalism to any EGUE(2) with $U(\Omega) \otimes SU(r)$ embedding and generated
by  random two-body  interaction with $SU(r)$ symmetry. Hereafter we call these
ensembles EGUE(2)-$SU(r)$ and they apply to both fermion and boson  systems. 

In Section 2, given is the general formulation based on Wigner-Racah algebra 
for lower order moments of the one- and two-point functions in eigenvalues 
generated by EGUE(2)-$SU(r)$  ($r$ is any positive integer,  $r \geq 1$).
Sections 3, 4 and 5 give analytical results for boson systems with  $r=1$, $r=2$
and $r=3$ respectively. In addition, some numerical results for lower order
correlations generated by these ensembles are also given in Section 5. Finally,
Section 6 gives concluding remarks.   

\section{EGUE(2)-$SU(r)$ ensembles: General formulation}

Consider a system of $m$ fermions or bosons in $\Omega$ number of sp levels each
$r$-fold degenerate. Then the SGA is $U(r\Omega)$ and it is possible to consider
$U(r\Omega) \supset U(\Omega) \otimes SU(r)$ algebra. Now, for random two-body
Hamiltonians preserving $SU(r)$ symmetry, one can introduce embedded GUE with
$U(\Omega) \otimes SU(r)$ embedding and this ensemble is called EGUE(2)-$SU(r)$.
Ensembles with $r=2,4$ for fermions correspond to fermions with spin and
spin-isospin $SU(4)$ symmetry. Similarly, for bosons $r=2,3$ are of interest.
Also $r=1$ gives back EGUE(2) and BEGUE(2) both. It is important to note that
the distinction between fermions and bosons is in the $U(\Omega)$ irreps that
need to be considered. Now we will give a formulation in terms of $SU(\Omega)$
Wigner-Racah algebra (the $SU(r)$ algebra involved will be simple as $H$ has
$SU(r)$ symmetry) that is valid for any $r$. The discussion in the  remaining
part of this Section is essentially from \cite{Ma-su4} but it is repeated
briefly not only for completeness but also to generalize it to any  $r$ and also
to bosons systems (in \cite{Ma-su4}, fermions with $r=4$ is used).

Let us begin with normalized two-particle states $\l.\l|f_2 F_2; v_2
\beta_2\r.\ran$ where the $U(r)$ irreps  $F_2=\{1^2\}$  and $\{2\}$  and the
corresponding $U(\Omega)$ irreps $f_2$ are $\{2\}$ (symmetric) and $\{1^2\}$
(antisymmetric) respectively for fermions and $\{1^2\}$ (antisymmetric) and
$\{2\}$ (symmetric) respectively for bosons.   Similarly $v_2$ are additional
quantum numbers that belong to $f_2$ and $\beta_2$ belong to $F_2$. As $f_2$
uniquely defines $F_2$, from now on we will drop $F_2$ unless it is explicitly
needed and also we will use the $f_2 \leftrightarrow F_2$ equivalence whenever
needed.  With $A^\dagger(f_2 v_2 \beta_2)$ and $A(f_2 v_2 \beta_2)$ denoting
creation and annihilation operators for the normalized two particle states, a
general two-body Hamiltonian operator $\wh$ preserving $SU(r)$ symmetry can be
written  as
\be
\wh= \wh_{\{2\}} + \wh_{\{1^2\}} =
\dis\sum_{f_2, v_2^i, v_2^f, \beta_2; f_2=\{2\}, \{1^2\}}
\;H_{f_2 v_2^i v_2^f}(2)\;
A^\dagger(f_2 v^f_2 \beta_2)\,A(f_2 v^i_2 \beta_2)\;.
\label{eq.egsu46}
\ee
In Eq. (\ref{eq.egsu46}), $H_{f_2 v_2^i v_2^f}(2)= \lan f_2 v^f_2 \beta_2 \mid H
\mid f_2 v^i_2 \beta_2\ran$ independent of the $\beta_2$'s. The uniform
summation over $\beta_2$ in  Eq. (\ref{eq.egsu46}) ensures that $\wh$ is $SU(r)$
scalar and therefore it will not connect  states with different $f_2$'s. 
However, $\wh$ is not a $SU(r)$ invariant operator.   Just as the two particle
states, we can denote the $m$ particle states by  $\l| f_m  v_m^f \beta_m^F
\ran$; $F_m=\widetilde{f}_m$ for fermions and $F_m=f_m$ for bosons.  Action of
$\wh$ on these states generates states that are degenerate with respect to
$\beta_m^F$ but not $v_m^f$. Therefore for a given $f_m$, there will be 
$d_\Omega(f_m)$  number of levels each with $d_r(\tilde{f}_m)$  number of
degenerate states.  Formula for the dimension $d_\Omega(f_m)$ is \cite{Wy-70}, 
\be
d_\Omega(f_m) = \dis\prod_{i<j=1}^\Omega \dis\frac{f_i-f_j+j-i}{j-i}\;,
\label{eq.egsu47}
\ee
where, $f_m=\{f_1,f_2,\ldots\}$. Equation (\ref{eq.egsu47}) also gives
$d_r(F_m)$  with the product ranging from $i=1$ to $r$ and replacing $f_i$ by
$F_i$. As $\wh$ is a $SU(r)$ scalar,  the $m$ particle $H$ matrix will be a
direct sum of matrices with each of them labeled by the $f_m$'s with dimension
$d_\Omega(f_m)$. Thus 
\be
H(m)=\dis\sum_{f_m}\;H_{f_m}(m) \oplus \;.
\label{eq.egsu48}
\ee
It should be noted that the  matrix elements of $H_{f_m}(m)$ matrices receive 
contributions from  both $H_{\{2\}}(2)$ and $H_{\{1^2\}}(2)$.

Embedded random matrix ensemble EGUE(2)-$SU(r)$ for a $m$ fermion or boson 
system with a fixed $f_m$, i.e. $\{H_{f_m}(m)\}$, is generated  by the ensemble
of $H$ operators given in  Eq. (\ref{eq.egsu46}) with  $H_{\{2\}}(2)$ and
$H_{\{1^2\}}(2)$ matrices replaced by independent GUE  ensembles of random
matrices,
\be
\{H(2)\} = \{H_{\{2\}}(2)\}_{GUE} \oplus \{H_{\{1^2\}}(2)\}_{GUE}\;.
\label{eq.egsu49}
\ee
In Eq. (\ref{eq.egsu49}), $\{--\}$ denotes ensemble.  Random variables defining
the real and imaginary parts of the matrix elements of $H_{f_2}(2)$ are 
independent Gaussian variables with zero center and variance given by (with bar
representing ensemble average),
\be
\overline{H_{f_2 v_2^1 v_2^2}(2)\;H_{f_2^\pr v_2^3 v_2^4}(2)}
= \delta_{f_2 f_2^\pr} \delta_{v_2^1 v_2^4} \delta_{v_2^2 v_2^3}\,
(\lambda_{f_2})^2\;.
\label{eq.egsu410}
\ee
Also, the independence of the $\{H_{\{2\}}(2)\}$ and $\{H_{\{1^2\}}(2)\}$ GUE 
ensembles imply,
\be
\overline{\l[H_{\{2\} v_2^1 v_2^2}(2)\r]^P\;
\l[H_{\{1^2\} v_2^3 v_2^4}(2)\r]^Q} 
= \l\{\;\overline{\l[H_{\{2\} v_2^1 v_2^2}(2)\r]^P}\;\r\}\;\;
\l\{\;\overline{\l[H_{\{1^2\} v_2^3 v_2^4}(2)\r]^Q}\;\r\}
\label{eq.egsu411}
\ee
for $P$ and $Q$ even and zero otherwise. Action of $\wh$ defined by Eq.
(\ref{eq.egsu46}) on $m$ particle basis states  with a fixed $f_m$, along with
Eqs. (\ref{eq.egsu410})-(\ref{eq.egsu411})  generates EGUE(2)-$SU(r)$ ensemble
$\{H_{f_m}(m)\}$; it is labeled by the  $U(\Omega)$ irrep $f_m$ with matrix
dimension $d_{\Omega}(f_m)$.

\begin{figure}[ht]
\centering
    \subfigure
    {
    \includegraphics[width=2.4in,height=3in]{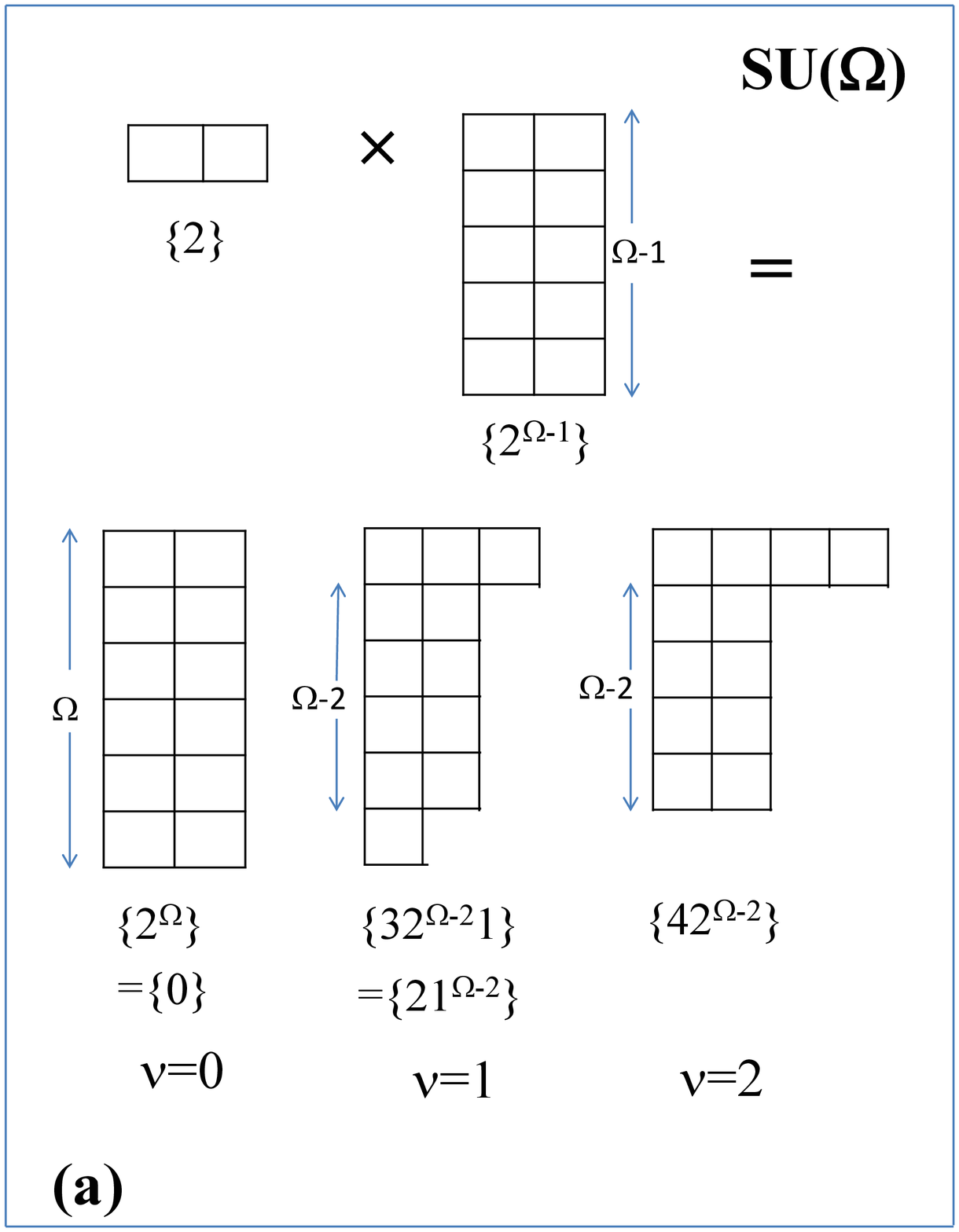}
        \label{young2a}
    }
    \subfigure
    {
        \includegraphics[width=2.4in,height=3in]{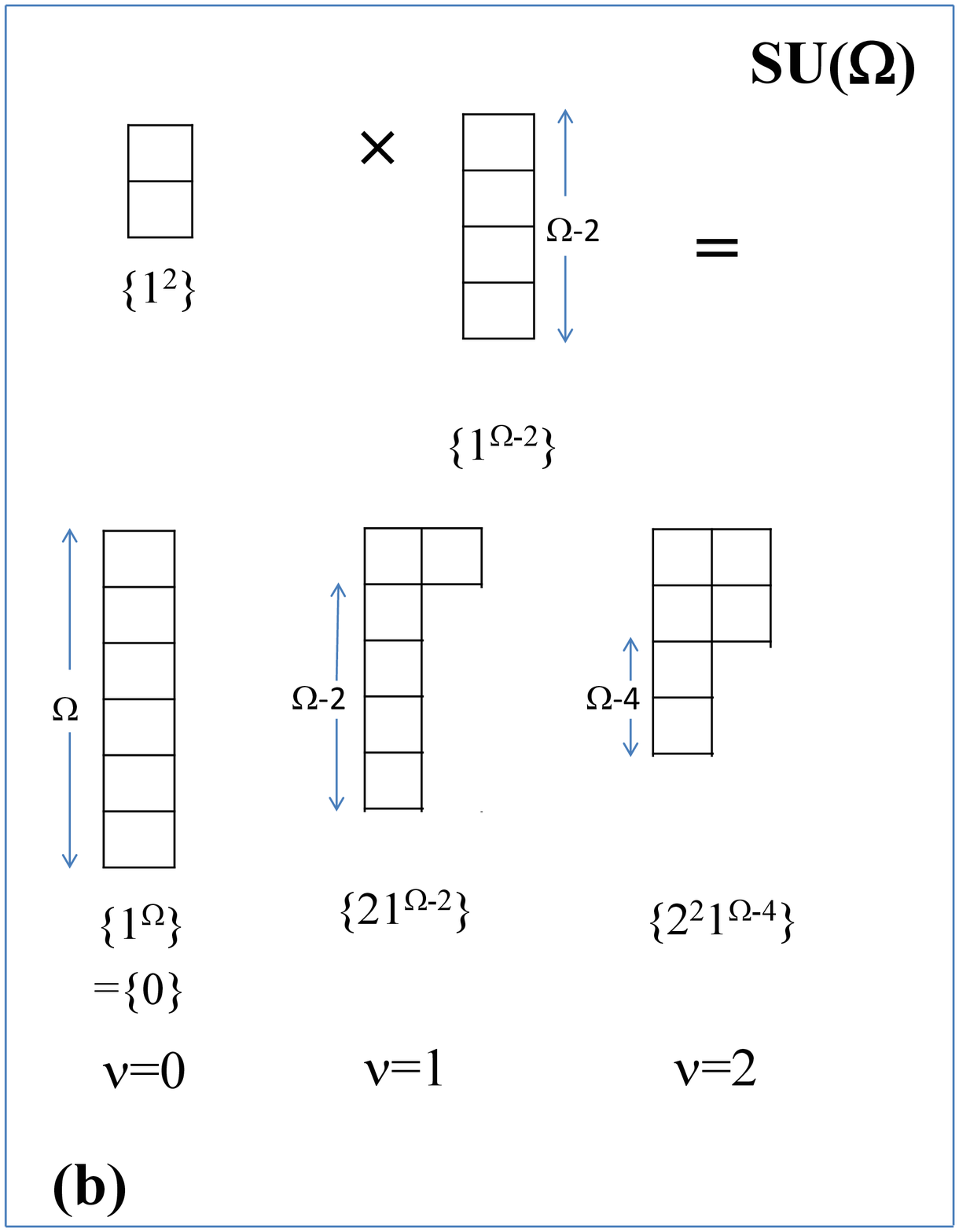}
        \label{young2b}
    }

\caption{Young tableaux for the tensorial parts of a two-body Hamiltonian with
respect to $SU(\Omega)$ algebra. Young tableaux for various (a)  tensorial parts
with respect to $SU(\Omega)$ for the $f_2=\{2\}$ part of $H$; (b) tensorial
parts with respect to $SU(\Omega)$ for the $f_2=\{1^2\}$ part of $H$.}

\label{young2}
\end{figure}

As shown in \cite{Ko-05,Ko-07,Ma-su4}, tensorial decomposition of $\wh$ with
respect to the embedding algebra $U(\Omega) \otimes SU(r)$ plays a crucial role
in generating analytical  results; as before  $U(\Omega)$ and  $SU(\Omega)$ are
used interchangeably. As $\wh$ preserves $SU(r)$, it transforms as the irrep
$\{0\}$ with respect to the $SU(r)$ algebra. However with respect to
$SU(\Omega)$, the tensorial characters, in Young tableaux notation, for
$f_2=\{2\}$ are $\bF_\nu=\{0\}$, $\{21^{\Omega-2}\}$ and $\{42^{\Omega-2}\}$
with $\nu=0,1$ and 2 respectively. Similarly for $f_2=\{1^2\}$ they are
$\bF_\nu= \{0\}$, $\{21^{\Omega-2}\}$ and $\{2^2 1^{\Omega-4}\}$ with
$\nu=0,1,2$ respectively. Note that $\bF_\nu=f_2 \times \overline{f_2}$ where
$\overline{f_2}$ is the irrep conjugate to $f_2$ and the $\times$ denotes
Kronecker product. Given a $U(\Omega)$ irrep
$\{f\}=\{f_1,f_2,\ldots,f_\Omega\}$, we have $\overline{f} = \{f_1-f_\Omega,
f_1-f_{\Omega-1}, \ldots,f_1-f_2,0\}$. Young tableaux for the  $F_\nu$'s are
shown in Fig. \ref{young2}.  Now, we can define unitary tensors $B$'s that are
scalars  in $SU(r)$ space,
\be
\barr{rcl}
B(f_2 \bF_\nu \omega_\nu) & = & \dis\sum_{v_2^i,v_2^f, \beta_2}\,
A^\dagger(f_2 v^f_2 \beta_2)\, A(f_2 v^i_2 \beta_2)\, \lan f_2
v_2^f\;\overline{f_2}\,\overline{v_2^i} \mid \bF_\nu \omega_\nu\ran
\\
& \times &
\lan F_2 \beta_2\;\overline{F_2}\,\overline{\beta_2} \mid 0 0\ran\;.
\earr \label{eq.egsu412}
\ee
In Eq. (\ref{eq.egsu412}), $\lan f_2 --- \ran$ are $SU(\Omega)$  Wigner 
coefficients  and $\lan F_2 --- \ran$ are $SU(r)$ Wigner coefficients. The 
expansion of $\wh$ in terms of $B$'s is,
\be
\wh=\dis\sum_{f_2, \bF_\nu, \;\omega_\nu}\;W(f_2 \bF_\nu \omega_\nu)\,
B(f_2 \bF_\nu \omega_\nu)\;.
\label{eq.egsu413}
\ee
The expansion coefficients $W$'s follow from the orthogonality of the
tensors $B$'s with respect to the traces over fixed $f_2$ spaces. Then we
have the most important relation needed for all the results given ahead,
\be
\overline{W(f_2 \bF_\nu \omega_\nu)W(f^\pr_2 \bF^\pr_\nu \omega^\pr_\nu)}
=\delta_{f_2 f^\pr_2} \delta_{\bF_\nu \bF^\pr_\nu} \delta_{\omega_\nu
\omega^\pr_\nu} \, (\lambda_{f_2})^2 d_r(F_2)\,.
\label{eq.egsu414}
\ee
This is derived starting with Eq. (\ref{eq.egsu413}) and using Eqs.
(\ref{eq.egsu49})-(\ref{eq.egsu412}) along with the sum rules for Wigner
coefficients appearing in Eq. (\ref{eq.egsu412}).

Turning to $m$ particle $H$ matrix elements, first we denote the $U(\Omega)$ and
$U(r)$ irreps by $f_m$ and $F_m$ respectively. Correlations generated by
EGUE(2)-$SU(r)$ between states with $(m,f_m)$ and $(m^\pr,f_{m^\pr})$ follow
from the covariance between the $m$-particle matrix elements of $H$. Now using
Eqs. (\ref{eq.egsu413}) and (\ref{eq.egsu414}) along with the Wigner-Eckart
theorem applied  using $SU(\Omega) \otimes SU(r)$ Wigner-Racah algebra (see for
example \cite{He-74a}) will give
\be
\barr{l}
\overline{H_{f_m v_m^i v_m^f}\,H_{f_{m^\pr} v_{m^\pr}^i v_{m^\pr}^f}}
\\ \\
=
\overline{\lan f_m F_m v_m^f \beta \mid H \mid f_m F_m v_m^i \beta\ran
\lan f_{m^\pr} F_{m^\pr} v_{m^\pr}^f \beta^\pr \mid H \mid f_{m^\pr}
F_{m^\pr} v_{m^\pr}^i \beta^\pr \ran }
\\ \\
= \dis\sum_{f_2, \bF_\nu,\; \omega_\nu} \; \dis\frac{(\lambda_{f_2})^2}
{d_{\Omega}(f_2)}\;
\dis\sum_{\rho,\rho^\pr}\; \lan f_m \mid\mid\mid B(f_2 \bF_\nu)
\mid\mid\mid
f_m\ran_\rho\;
\lan f_{m^\pr} \mid\mid\mid B(f_2 \bF_\nu) \mid\mid\mid f_{m^\pr}
\ran_{\rho^\pr} \\ \\
\times \lan f_m v_m^i\;\bF_\nu \omega_\nu
\mid f_m v_m^f\ran_\rho\;
\lan f_{m^\pr} v_{m^\pr}^i\;\bF_\nu \omega_\nu \mid f_{m^\pr}
v_{m^\pr}^f\ran_{\rho^\pr}\,;
\\ \\
\lan f_m \mid\mid\mid
B(f_2 \bF_\nu) \mid\mid\mid f_m\ran_\rho\,=
\dis\sum_{f_{m-2}}\;F(m)\, \dis\frac{\cn_{f_{m-2}}}{\cn_{f_m}} \;
\dis\frac{U(f_m \overline{f_2} f_m f_2; f_{m-2} \bF_\nu)_\rho}{U(f_m
\overline{f_2} f_m f_2; f_{m-2} \{0\})} \;.
\earr \label{eq.egsu415}
\ee
Here the summation in the last equality is over the multiplicity index $\rho$
and this arises  as  $f_m \times \bF_\nu$ gives in general more than once the
irrep $f_m$. In Eq. (\ref{eq.egsu415}), 
\be
F(m)=-m(m-1)/2\;,
\label{eq.egsu-aa}
\ee  
$d_{\Omega}(f_m)$ is given by Eq. (\ref{eq.egsu47}) and  $\lan \ldots \ran$ and
$U(\ldots)$ are $SU(\Omega)$ Wigner and Racah coefficients respectively.
Similarly, $\cn_{f_m}$ is  dimension with respect to the  $S_m$ group
\cite{Wy-70},
\be
\cn_{f_m} = \dis\frac{m!\dis\prod_{i<k=1}^p (\ell_i-\ell_k)}
{\ell_1!\;\ell_2!\ldots\ell_p!}\;;\;\;\;\;\ell_i=f_i+p-i \;.
\label{eq.egsu4-n1}
\ee
Note that $p$ denotes total number of rows in the Young tableaux for $f_m$.
 
Lower order cross correlations between states with different $(m,f_m)$ are given
by the  normalized bivariate moments $\Sigma_{PQ}\l(m,f_m: m^\pr ,f_{m^\pr}\r)$,
$P=Q=1,2$ of the two-point function $S^\rho$ where,  with $\rho^{m,f_m}(E)$
defining fixed-$(m,f_m)$ density of states,
\be
\barr{l}
S^{m f_m:m^\pr f_{m^\pr}}(E,E^\pr) = \overline{\rho^{m,f_m}(E)
\rho^{m^\pr ,
f_{m^\pr}}(E^\pr)} - \overline{\rho^{m,f_m}(E)}\;\;\overline{\rho^{m^\pr ,
f_{m^\pr}}(E^\pr)}\;\;; \\ \\
\Sigma_{11}\l( {m,f_m}: {m^\pr ,f_{m^\pr}}\r) = \overline{
\lan H \ran^{m,f_m} \; \lan H \ran^{m^\pr,f_{m^\pr}}}/\dis\sqrt{\,
\overline{\lan H^2 \ran^{m,f_m}} \; \overline{\lan H^2 \ran^{m^\pr,
f_{m^\pr}}}}\;, \\ \\
{\Sigma}_{22}\l( {m,f_m}: {m^\pr ,f_{m^\pr}}\r) = \overline{
\lan H^2 \ran^{m,f_m} \; \lan H^2 \ran^{m^\pr,f_{m^\pr}}}/\l[\,
\overline{\lan H^2 \ran^{m,f_m}} \; \overline{\lan H^2 \ran^{m^\pr,
f_{m^\pr}}}\,\r] -1\;.
\earr \label{eq.egsu416}
\ee
In Eq. (\ref{eq.egsu416}), $\overline{\lan H^2 \ran^{m,f_m}}$ is the second
moment (or variance) of the eigen value density  $\overline{\rho^{m,f_m}(E)}$ 
and  its centroid $\overline{\lan H\ran^{m,f_m}}=0$ by definition. As $\lan H
\ran^{m,f_m}$ is the trace of $H$  (divided by dimensionality)  in $(m,f_m)$
space, only $\bF_\nu = \{0\}$ will generate $\overline{\lan H \ran^{m,f_m} \; 
\lan H \ran^{m^\pr,f_{m^\pr}}}$. Then trivially,
\be
\barr{rcl}
\overline{\lan H \ran^{m,f_m} \; \lan H \ran^{m',f_{m^\pr}}}
& = & \dis\sum_{f_2} \dis\frac{\l(\lambda_{f_2}\r)^2}{d_{\Omega}(f_2)}\;P^{f_2}
(m,f_m)\;P^{f_2}(m^\pr,f_{m^\pr})\;;\\ \\
P^{f_2}(m,f_m) & = & F(m)\dis\sum_{f_{m-2}}\;\dis\frac{\cn_{f_{m-2}}}{
\cn_{f_m}}\;. 
\earr \label{eq.egsu417} 
\ee 
Writing $\overline{\lan H^2 \ran^{m,f_m}}$ explicitly in terms  of $m$ particle
$H$ matrix elements, 
$$
\overline{\lan H^2 \ran^{m,f_m}} =
[d(f_m)]^{-1}\sum_{v_m^1 ,  v_m^2} \, \overline{H_{f_m v_m^1 v_m^2}\,H_{f_m
v_m^2 v_m^1}}\;,
$$
and applying Eq. (\ref{eq.egsu415}) and the orthonormal properties of the 
$SU(\Omega)$ Wigner coefficients lead to
\be
\overline{\lan H^2 \ran^{m,f_m}} = \dis\sum_{f_2}
\dis\frac{(\lambda_{f_2})^2}{d_{\Omega}(f_2)}
\dis\sum_{\nu=0,1,2}\cq^\nu(f_2:m,f_m)
\label{eq.egsu418}
\ee
where
\be
\cq^\nu(f_2:m,f_m)=
\l[F(m)\r]^2 \dis\sum_{f_{m-2},f^\pr_{m-2}}
\dis\frac{\cn_{f_{m-2}}}{\cn_{f_m}}\dis\frac
{\cn_{f^\pr_{m-2}}}{\cn_{f_m}}X_{UU}(f_2;f_{m-2},f^\pr_{m-2};\bF_\nu)\;.
\label{eq.egsu419}
\ee
The $X_{UU}$ function involves $SU(\Omega)$ Racah coefficients,
\be
\barr{l}
X_{UU}(f_2;f_{m-2},f^\pr_{m-2};\bF_\nu) = \\
\dis\sum_{\rho}\dis\frac
{U(f_m,\overline{f_2},f_m,f_2;
f_{m-2},\bF_\nu)_\rho U(f_m,\overline{f_2},f_m,f_2;f^\pr_{m-2},\bF_\nu)
_\rho}{U(f_m,\overline{f_2},f_m,f_2;f_{m-2},\{0\})
U(f_m,\overline{f_2},f_m,f_2;f^\pr_{m-2},\{0\})}\;.
\earr \label{eq.egesu4-n2}
\ee
Summation over the multiplicity index $\rho$ in Eq. (\ref{eq.egesu4-n2}) arises
naturally in applications to physical problems as all the physically relevant
results should be independent of $\rho$ which is a label for equivalent
$SU(\Omega)$ irreps. It is easy to see that, 
\be
\cq^{\nu=0}(f_2:m,f_m) = \l[P^{f_2}(m,f_m) \r]^2\;.
\label{eq.egsu4-n3}
\ee 
Eqs. (\ref{eq.egsu417})-(\ref{eq.egsu419}) and Table 4 of \cite{Ma-su4}  will
allow us to calculate covariances ${\Sigma}_{11}$ in energy centroids; Table 4
of \cite{Ma-su4} is a simplified version of the tables in \cite{He-74}. For the
covariances ${\Sigma}_{22}$ in  spectral variances, the formula is \cite{Ma-su4}
\be
\barr{l}
{\Sigma}_{22}(m,f_m;m^\pr,f_{m^\pr}) = \dis\frac{X_{\{2\}}+
X_{\{1^2\}} +4X_{\{1^2\}\{2\}}}{\overline{\lan H^2\ran^{m,f_m}}\;
\overline{\lan H^2\ran^{m^\pr,f_{m^\pr}}}}\;;\\
X_{f_2} = \dis\frac{2(\lambda_{f_2})^4}{\l[d_{\Omega}(f_2)\r]^2}
\dis\sum_{\nu=0,1,2} \l[ d_{\Omega}(\bF_\nu)\r]^{-1}\cq^\nu(f_2:m,f_m)
\cq^\nu(f_2:m^\pr,f_{m^\pr})\;, \\
X_{\{1^2\}\{2\}} =
\dis\frac{\lambda^2_{\{2\}}\lambda^2_{\{1^2\}}}{d(\{2\})d(\{1^2\})}
\dis\sum_{\nu=0,1} \l[ d_{\Omega}(\bF_\nu)\r]^{-1} \car^\nu(m,f_m)\;
\car^\nu(m^\pr,f_{m^\pr})\;. 
\earr \label{eq.egsu420}
\ee
Here $d_{\Omega}(\bF_\nu)$ is the dimension of the irrep $\bF_\nu$, and we have
$d_{\Omega}(\{0\}) = 1$, $d_{\Omega}(\{2,1^{\Omega-2}\}) = \Omega^2-1$, 
$d_{\Omega}(\{4,2^{\Omega-2}\}) = \Omega^2(\Omega+3)(\Omega-1)/4$, and 
$d_{\Omega}(\{2^2,1^{\Omega-4}\}) = \Omega^2(\Omega-3)(\Omega+1)/4$. Note that 
$\cq^\nu(f_2:m,f_m)$ are  defined in Eq. (\ref{eq.egsu419}). The functions
$R^\nu(m,f_m)$ also involve $SU(\Omega)$ $U$-coefficients,
\be
\barr{l}
R^\nu(m,f_m) = \l[ F(m) \r]^2 \dis\sum_{f_{m-2},f^\pr_{m-2}}
\dis\frac{\cn_{f_{m-2}}}{\cn_{f_m}}\dis\frac{\cn_{f_{m-2}^\pr}}{\cn_{f_m}}
Y_{UU}(f_{m-2},f^\pr_{m-2};\bF_\nu)\;; \\
Y_{UU}(f_{m-2},f^\pr_{m-2};\bF_\nu) = \\
\dis\sum_{\rho}
\dis\frac{U(f_m,\{1^{\Omega-2}\},f_m,\{1^2\};f_{m-2},\bF_\nu)
_\rho\,U(f_m,\{2^{\Omega-1}\},f_m,\{2\};
f_{m-2}^\pr,\bF_\nu)_\rho}{U(f_m,\{1^{\Omega-2}\},f_m,\{1^2\};f_{m-2},
\{0\})\,U(f_m,\{2^{\Omega-1}\},f_m,\{2\};f_{m-2}^\pr,\{0\})}\;.
\earr \label{eq.egsu4-n4}
\ee
In $Y_{UU}(f_{m-2},f^\pr_{m-2};\bF_\nu)$, $f_{m-2}$ comes from $f_m \otimes
\{1^{\Omega-2}\}$ and $f^\pr_{m-2}$ comes  from $f_m \otimes \{2^{\Omega-1}\}$.
Similarly, the summation is over $\nu=0$ and $1$ only as $\nu=2$ parts for
$f_2=\{2\}$ and $\{1^2\}$ are different. Formulas for $Y_{UU}$ are given in
Table 7 of \cite{Ma-su4} and they are simplified version of the results in
\cite{He-74}. It is useful to note that,
\be
\car^{\nu=0}(m,f_m) = P^{\{2\}}(m,f_m)P^{\{1^2\}}(m,f_m)\;.
\label{eq.egsu4-n5}
\ee 
Compact analytical results collected in  Tables 4 and 7 of \cite{Ma-su4} for 
$X_{UU}$ and $Y_{UU}$ and Eqs. (\ref{eq.egsu47}), (\ref{eq.egsu4-n1}) - (
\ref{eq.egsu4-n5}) will allow one to derive analytical/numerical results for
spectral variances and covariances in energy centroids and variances for any 
EGUE(2)-$SU(r)$ for fermion or boson systems. 

\begin{figure}[ht]
\centering
    
\includegraphics[width=3in,height=4in]{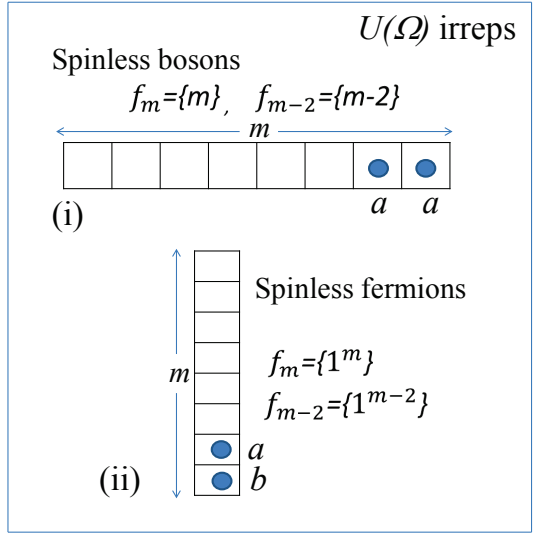}
    
\caption{Young tableaux denoting the $SU(\Omega)$ irreps $f_m=\{m\}$ and 
$\{1^m\}$ as appropriate for (i) spinless boson and (ii) spinless fermion 
systems. Removal of two boxes generating $m-2$ particle irreps $f_{m-2}$ for 
these systems are also shown in the figure. For (i) only the irrep $f_2=\{2\}$ 
will apply and similarly for (ii) only $\{1^2\}$ will apply.}

\label{young-fermi-bose}
\end{figure}

\section{Results for BEGUE(2): $r=1$}

Simplest of the EGUE(2)-$SU(r)$ are the EGUEs with $r=1$ and they corresponds to
EGUE(2) and BEGUE(2) depending on totally antisymmetric or symmetric $f_m$ one
considers. Also they correspond to $k=2$ in \cite{Ko-05} and \cite{Asa-02} for
fermion and boson systems respectively. As detailed results for fermion systems
are available in \cite{Ko-05,Ko-07,Ma-12}, in the present Section and in
the next two Sections
we consider only boson systems. Let us begin with BEGUE(2). For this
ensemble, in order to apply the formulas given Section 2 for $\lan H^2\ran$, 
$\Sigma_{11}$ and $\Sigma_{22}$, first we need formulas for $X_{UU}$ and
$Y_{UU}$. Some of these, taken from Tables 4 and 7 of \cite{Ma-su4}, are given
in Table \ref{xuuyuu} by reducing them to much small number of formulas. For
applying  these formulas, we need the 'axial distances' $\tau_{ij}$ for the
boxes $i$ and $j$ in a given  Young tableaux. Given a
$f_m=\{f_1,f_2,\ldots,f_\Omega\}$ we have,
\be
\tau_{ij} = f_i-f_j+j-i\;.
\label{eq.begue2-1}
\ee
In terms of $\tau_{ij}$ the functions $\Pi_a^{(b)}$, $\Pi_b^{(a)}$, 
$\Pi_a^{(bc)}$, $\Pi_a^{\pr}$ and $\Pi_a^{\pr \pr}$ are defined as,
\be
\barr{rcl}
\Pi_a^{(b)} & = & \dis\prod_{i=1,2,\ldots,\Omega;i\neq a,i \neq b}\;
\l( 1-1/\tau_{ai}\r)\;, \\ 
\\
\Pi_b^{(a)} & = & \dis\prod_{i=1,2,\ldots,\Omega;i\neq a,i \neq b}\;
\l( 1-1/\tau_{bi}\r)\;,\\
\\
\Pi_a^{(bc)} & = & \dis\prod_{i=1,2,\ldots,\Omega;i\neq a,i \neq b,
i \neq c}\; \l( 1-1/\tau_{ai}\r) \;;\;\;a \neq b \neq c\;,\\
\\
\Pi^{\pr}_a & = & \dis\prod_{i=1,2,\ldots,\Omega;i\neq a}\;
\l( 1-1/\tau_{ai}\r) \;,\\ 
\\
\Pi^{\pr\pr}_a & = & \dis\prod_{i=1,2,\ldots,\Omega;i\neq a}\;
(1-2/\tau_{ai})\;.
\earr \label{eq.begue2-2}
\ee
With these we can calculate $X_{UU}$ and $Y_{UU}$; see \cite{Ma-su4} for full
discussion. For BEGUE(2), the algebra $U(\Omega) \otimes SU(r)$ with $r=1$
reduces to just $U(\Omega)$ or $SU(\Omega)$. Similarly, $f_m$ is the totally
symmetric irrep $\{m\}$ and $f_{m-2}=\{m-2\}$. Therefore to generate $f_{m-2}$
only the action of removal of $\{2\}$ from $f_m$ is allowed. Denoting the last
two boxes of $f_m$ by $a$ and $a$ (note that we can remove only boxes from the
right end to get a proper Young Tableaux and also boxes in a given row must have
the same symbol to apply the results in Table \ref{xuuyuu}) as shown in Fig.
\ref{young-fermi-bose}, we have
\be
\barr{rcl}
\tau_{ai} & = & m+i-1\;,\\
\Pi_a^{\pr} & = & \dis\frac{m}{m+\Omega-1} \;,\\
\\
\Pi_a^{\pr \pr} & = & \dis\frac{m(m-1)}{(m+\Omega-1)(m+\Omega-2)}\;.
\earr \label{eq.begue2-3}
\ee
Similarly $\cn_{f_m}=1$ and $\cn_{f_{m-2}}=1$ as both are symmetric irreps. Now
the formulas in Table \ref{xuuyuu} will give $X_{UU}$ and then using
Eq. (\ref{eq.egsu419}) we have,
\be
\barr{rcl}
\cq^{\nu=0}(\{2\};m,\{m\}) & = & \dis\frac{m^2 (m-1)^2}{4}\;,\\ \\
\cq^{\nu=1}(\{2\};m,\{m\}) & = & \dis\frac{m^2 (m-1)^2}{4} \dis\frac{2 
(\Omega+m)(\Omega^2-1)}{m (\Omega+2)}\;,\\ \\
\cq^{\nu=2}(\{2\};m,\{m\}) & = & \dis\frac{m^2 (m-1)^2}{4} \dis\frac{ 
\Omega^2(\Omega-1)(\Omega+m)(\Omega+m+1)}{2(\Omega+2)m(m-1)}\;.
\earr \label{eq.begue2-4}
\ee
These and Eq. (\ref{eq.egsu418}) will give,
\be
\barr{rcl}
\lan H^2 \ran^{\{m\}} & = & \lambda_{\{2\}}^2\;\dis\binom{m}{2}
\dis\binom{\Omega+m-1}{2} = \lambda_{\{2\}}^2\; \Lambda^{\nu=0}(\Omega,m,2) \;;\\ \\
\Lambda^\nu(\Omega,m,k) & = & \dis\binom{m-\nu}{k}\;\dis\binom{\Omega+m+\nu-1}{k} \;. 
\earr \label{eq.begue2-5}
\ee
This agrees with the result stated in \cite{Asa-02,Ko-05}. As $P^{\{2\}}(m,
\{m\}) =-m(m-1)/2$, we have easily,
\be
{\hat{\Sigma}}_{11}(\{m\},\{m^\pr\}) = \dis\frac{2\dis\sqrt{m(m-1)(m^\pr)(
m^\pr-1)}}{\Omega(\Omega+1)\dis\sqrt{(\Omega+m-1)(\Omega+m-2)(\Omega+m^\pr-1)
(\Omega+m^\pr-2)}}\;.
\label{eq.begue2-6}
\ee
Again this agrees, for $m=m^\pr$ with the result stated in \cite{Asa-02,Ko-05}. Further,  
${\hat{\Sigma}}_{22}$ is determined only by $X_{\{2\}}$ defined in Eq.
(\ref{eq.egsu420}) and then, using Eq. (\ref{eq.begue2-4}), we have
\be
\barr{l}
{\hat{\Sigma}}_{22}(\{m\},\{m^\pr\}) = \dis\frac{2}{36\dis\binom{\Omega+2}{3}^2 
(\Omega+3)\dis\binom {\Omega+m-1}{2}\dis\binom{\Omega+m^\pr -1}{2}} \\
\\
\times \l[4\Omega^2(\Omega-1)\dis\binom{\Omega+m+1}{2} \dis\binom{\Omega+m^\pr +1}{2}
+ 4(\Omega+2)^2(\Omega+3) \dis\binom{m}{2}\dis\binom{m^\pr}{2}\r. \\
\\
\l.+4(\Omega^2-1)(\Omega+3) (m-1)(\Omega+m) (m^\pr -1)(\Omega+m^\pr)\r]\;.
\earr \label{eq.begue2-7}
\ee 
For $m=m^\pr$, it can be verified that Eq. (\ref{eq.begue2-7}) reduces to
\be
{\hat{\Sigma}}_{22}(\{m\},\{m^\pr\}) = \dis\frac{2}{\l(\Omega_m\r)^2} 
\dis\sum_{\nu=0}^2\;\dis\frac{\l[\Lambda^{\nu}(\Omega,m,m-2)\r]^2 
d_{\Omega}(\bF_\nu)}{\l[\Lambda^{\nu^\pr=0}(\Omega,m,2)\r]^2}\;;\;\;\;
\Omega_m = {\Omega+m-1 \choose m}
\label{eq.begue2-8}
\ee
and this agrees with the result given in \cite{Asa-02}. Note that $\bF_\nu$ is
$\{0\}$, $\{21^{\Omega-2}\}$ and $\{42^{\Omega-2}\}$ for $\nu=0$, $1$ and $2$
respectively. It is useful to mention that Eqs. (\ref{eq.begue2-6}) and 
(\ref{eq.begue2-7}) follow from the results for fermion systems given in 
\cite{Ko-06} with $\Omega \rightarrow -\Omega$ symmetry. Finally, it is useful 
to  mention that in the $m \rightarrow \infty$ and $\Omega$ finite limit we 
have,
\be
\barr{l}
{\hat{\Sigma}}_{11}(\{m\},\{m^\pr\}) = \dis\frac{2}{\Omega (\Omega+1)}\;,\\
\\
{\hat{\Sigma}}_{22}(\{m\},\{m^\pr\}) = 8\dis\frac{\Omega^2(\Omega-1) +
(\Omega+2)^2(\Omega+3)+4(\Omega^2-1)(\Omega+3)}{\Omega^2 (\Omega+1)^2 
(\Omega+2)^2 (\Omega+3)}\;.
\earr \label{eq.begue2-9}
\ee
Non-vanishing of ${\hat{\Sigma}}_{11}$ and ${\hat{\Sigma}}_{22}$ for finite 
$\Omega$ in the $m \rightarrow \infty$ limit is interpreted in 
\cite{Asa-01,Asa-02}  as non-ergodicity of BEGUE ensembles. See the discussion 
in \cite{Ch-03} for the resolution of this problem.

\begin{table}[ht]
\caption{Formulas for $X_{UU}(f_2;f_{m-2},f^\pr_{m-2};\bF_\nu)$ and 
$Y_{UU}(f_{m-2},f^\pr_{m-2};\bF_\nu)$ with $\nu=1,2$.}
\begin{center}
\begin{tabular}{lc}
\hline \hline
$\{f_{m-2}\}\,\{f^\pr_{m-2}\}$ & $X_{UU} (\{1^2\}; f_{m-2}, 
f^\pr_{m-2}; \{2^\nu,1^{\Omega-2\nu}\})$ \\ 
\hline \hline
$\{f(ab)\}\,\{f(ab)\}$ & $\dis\frac{\Omega}{(\Omega-2)} 
\l\{\delta_{\nu,2}+\dis\frac{(\Omega-1)(\Omega-2)}{2\Pi_a^{(b)}\Pi_b^{(a)}}\,
\delta_{\nu,2}
+(3-2\nu)\,\dis\frac{(\Omega-1)}{2}\r.$ \\ 
& $\l. \times 
\l[ \l(1+\dis\frac{1}{\tau_{ab}}\r)\dis\frac{1}{\Pi_b^{(a)}}+
\l(1-\dis\frac{1}{\tau_{ab}}\r)\dis\frac{1}{\Pi_a^{(b)}}-\dis\frac{4}{\Omega}
\delta_{\nu,1}\r]\r\}$ \\ 
$\{f(ab)\}\,\{f(ac)\}$ & $\dis\frac{\Omega(\Omega-1)}{2(\Omega-2)}
\l\{\dis\frac{2}{(\Omega-1)}\delta_{\nu,2}-\dis\frac{4}{\Omega}\delta_{\nu,1}+
(3-2\nu)\,\dis\frac{1}{\Pi_a^{(bc)}}\r\}$ \\ 
\hline \hline
$\{f_{m-2}\}\,\{f^\pr_{m-2}\}$ & $X_{UU} (\{2\}; f_{m-2}, 
f^\pr_{m-2}; \{2\nu,\nu^{\Omega-2}\})$ \\ 
\hline \hline
$\{f(ab)\}\,\{f(ab)\}$ & $\dis\frac{\Omega(\Omega+1)}{2} 
\l\{\dis\frac{1}{\Pi_a^{(b)}\Pi_b^{(a)}}\delta_{\nu,2}+\dis\frac{2}{
(\Omega+1)(\Omega+2)}\delta_{\nu,2}
\r.$ \\ 
& $\l.+(3-2\nu)\dis\frac{1}{(\Omega+2)}\l[\dis\frac{(\tau_{ab}-1)^2}{\tau_{ab}
(\tau_{ab}+1)}\dis\frac{1}{\Pi_b^{(a)}} + \dis\frac{(\tau_{ab}+1)^2}
{\tau_{ab}(\tau_{ab}-1)}\dis\frac{1}{\Pi_a^{(b)}} -\dis\frac{4}{\Omega}
\delta_{\nu,1}\r]\r\}$ \\ 
$\{f(aa)\}\,\{f(aa)\}$ & $\dis\frac{\Omega}{(\Omega+2)} 
\l\{\delta_{\nu,2}+(3-2\nu)\dis\frac{2(\Omega+1)}{\Pi_a^{\pr}}+\dis\frac{
(\Omega+1)(\Omega+2)}
{2\,\Pi_a^{\pr\pr}}\delta_{\nu,2}-\dis\frac{2(\Omega+1)}{
\Omega}\delta_{\nu,1}\r\}$ \\ 
$\{f(aa)\}\,\{f(bb)\}$ & $-\dis\frac{2(\Omega+1)}{(\Omega+1)}\delta_{\nu,1} 
+ \dis\frac{\Omega}{(\Omega+2)}\delta_{\nu,2}$ \\
$\{f(aa)\}\,\{f(ab)\}$ & $\dis\frac{\Omega}{(\Omega+2)}
\l\{\delta_{\nu,2}+(3-2\nu)\dis\frac{(\Omega+1)(\tau_{ab}+1)}{(\tau_{ab}-1)
\Pi_a^{(b)}}-\dis\frac{2(\Omega+1)}{\Omega}\delta_{\nu,1}\r\}$ \\
\hline \hline
$\{f_{m-2}\}\,\{f^\pr_{m-2}\}$ &  $Y_{UU} (f_{m-2}, 
f^\pr_{m-2}; \{2,1^{\Omega-2}\})$ \\ 
\hline \hline
$\{f(ab)\}\,\{f(ab)\}$ & $-\dis\frac{\Omega}{2}
\l[\dis\frac{(\Omega^2-1)}{(\Omega^2-4)}\r]^{1/2} 
\l\{\l(1+\dis\frac{1}{\tau_{ab}}\r)\dis\frac{1}{\Pi_a^{(b)}} +
\l(1-\dis\frac{1}{\tau_{ab}}\r)\dis\frac{1}{\Pi_b^{(a)}}-
\dis\frac{4}{\Omega}\r\}$ \\ 
$\{f(ab)\}\,\{f(ac)\}$ & $-\dis\frac{\Omega}{2}
\l[\dis\frac{(\Omega^2-1)}{(\Omega^2-4)}\r]^{1/2} \l\{ 
\l(1+\dis\frac{1}{\tau_{ac}}\r)\dis\frac{1}{\Pi_a^{(b)}}-
\dis\frac{4}{\Omega}\r\}$ \\ 
$\{f(ab)\}\,\{f(aa)\}$ & $-\Omega
\l[\dis\frac{(\Omega^2-1)}{(\Omega^2-4)}\r]^{1/2} \l\{ 
\dis\frac{1}{\Pi_a^{(b)}}-\dis\frac{2}{\Omega}\r\}$ \\ 
\hline\hline
\end{tabular}
\end{center}
\label{xuuyuu}
\end{table}

\section{Embedded Gaussian Unitary Ensemble for bosons with $F$-spin: 
BEGUE(2)-SU(2) with $r=2$}

\begin{figure}[ht]
\centering
    
\includegraphics[width=3in,height=4in]{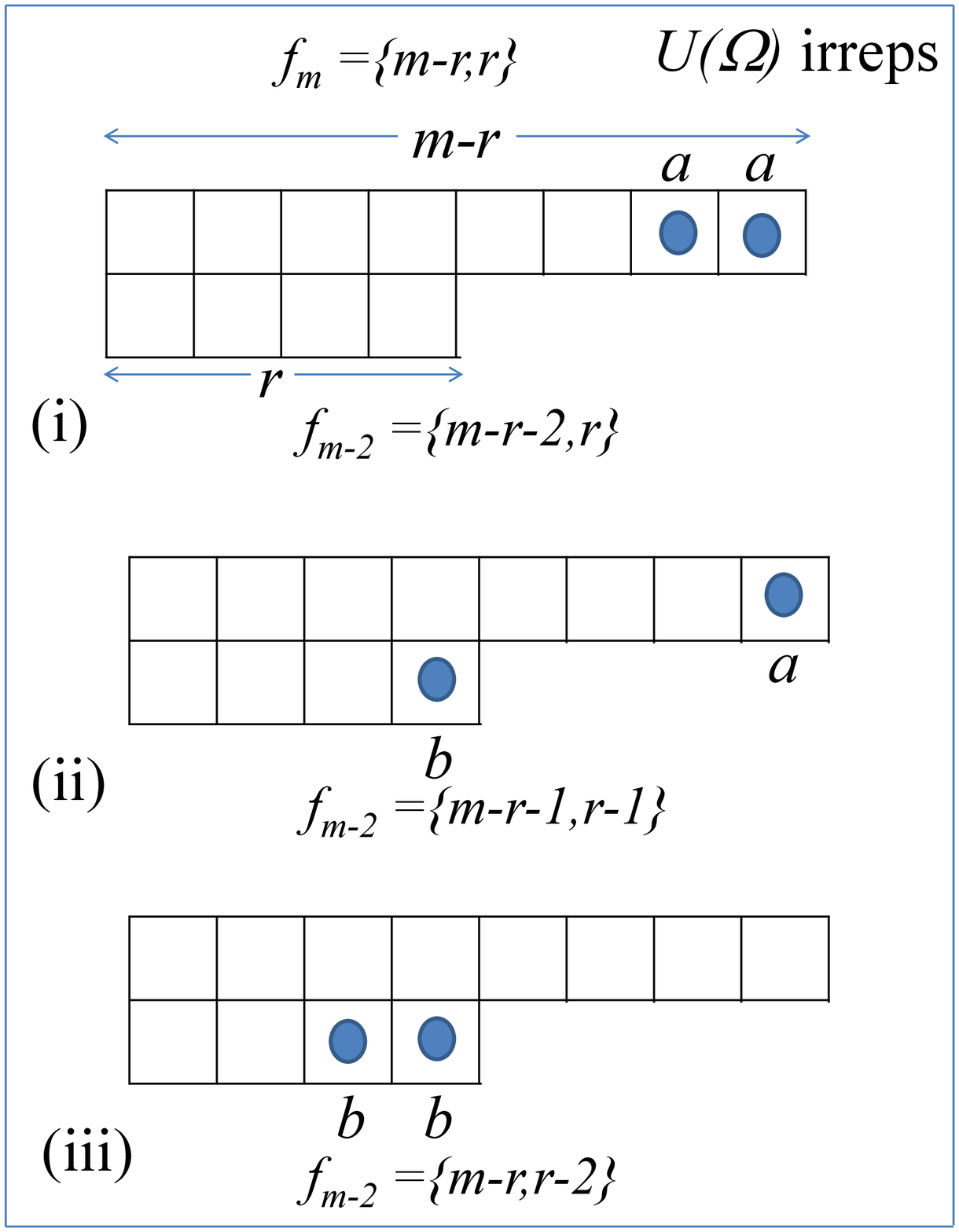}
    
\caption{Young tableaux denoting the two-rowed $SU(\Omega)$ irreps $f_m = \{m-r,
r\}$ appropriate for BEGUE(2)-$SU(2)$. Removal of two boxes generating $m-2$ 
particle irreps $f_{m-2}$ are also shown in the figure. For (ii) both the irreps
$f_2=\{2\}$ and $\{1^2\}$ will apply while for (ii) and (iii) only $\{2\}$ will
apply.}

\label{young-su2-bose}
\end{figure}

For two species boson systems we have BEGUE(2)-$SU(2)$ and then the formulation
in Section 2 with $r=2$ will be applicable. Here the two species are assumed to
be the two components of a fictitious $F$-spin as discussed recently in
\cite{Ma-12}. For such a $m$ boson system, the $SU(\Omega)$ irreps will be two
rowed denoted by $f_m=\{m-r,r\}$ with $F=\frac{m}{2}-r$. With this, there are
three allowed $f_{m-2}$ irreps as shown in Fig. \ref{young-su2-bose}. The irreps
in (i) and (iii) in the figure can be obtained by removing $f_2=\{2\}$ from
$f_m$.  However for (ii) in the figure both $\{2\}$ and $\{1^2\}$ will apply.
For $f_{m-2}=\{m-r-2,r\}$ irrep [this corresponds to (i) in Fig.
\ref{young-su2-bose}] we have
\be
\barr{rcl}
\tau_{a2} & = & m-2r+1\;,\\
\tau_{ai} & = & m-r+i-1\;;\; i=3,4,\ldots,\Omega\;, \\
\Pi_a^\pr & = & \dis\frac{(m-2r)(m-r+1)}{(m-2r+1)(m-r+\Omega -1)}\;, \\
\Pi_a^{\pr \pr} & = & \dis\frac{(m-2r-1)(m-r)(m-r+1)}{(m-2r+1)(m-r+\Omega -1)
(m-r+\Omega-2)} \;.
\earr \label{eq.bege-su2-1}
\ee
Similarly for $f_{m-2}=\{m-r,r-2\}$ irrep [this corresponds to (iii) in Fig. 
\ref{young-su2-bose}] we have
\be
\barr{rcl}
\tau_{b1} & = & 2r-m-1\;,\\
\tau_{bi} & = & r+i-2\;,\;i=3,4,\ldots,\Omega\\
\Pi_b^\pr & = & \dis\frac{(r)(2r-m-2)}{(2r-m-1)(r+\Omega -2)}\;, \\
\Pi_b^{\pr \pr} & = & \dis\frac{(2r-m-3)(r)(r-1)}{(2r-m-1)(r+\Omega -2)(r+
\Omega-3)} \;.
\earr \label{eq.bege-su2-2}
\ee
Finally, for $f_{m-2}=\{m-r-1,r-1\}$ irrep [this corresponds to (ii) in Fig. 
\ref{young-su2-bose}] we have
\be
\barr{rcl}
\tau_{ab} & = & m-2r+1=2F+1\;,\\
\tau_{ai} & = & m-r+i-1,\;\;\;\tau_{bi}=r+i-2\;;\;\;i=3,4,\ldots,\Omega,\\
\Pi_a^{(b)} & = & \dis\frac{(m-r+1)}{(m-r+\Omega -1)}\;, \\
\Pi_b^{(a)} & = & \dis\frac{(r)}{(r+\Omega -2)} \;.
\earr \label{eq.bege-su2-3}
\ee
These and $\cn_{f_{m-2}}/\cn_{f_m}$ will give the formulas for the lower order
moments of one and two point functions as described in Section 2. The
dimension ratios needed are,
\be
\barr{rcl}
\dis\frac{\cn_{\{m-r-2,r\}}}{\cn_{\{m-r,r\}}} & = & \dis\frac{(m-r)(m-r+1)
(m-2r-1)}{m(m-1)(m-2r+1)}\;,\\
\\
\dis\frac{\cn_{\{m-r-1,r-1\}}}{\cn_{\{m-r,r\}}} & = & \dis\frac{r(m-r+1)}{m
(m-1)}\;,\\
\\
\dis\frac{\cn_{\{m-r,r-2\}}}{\cn_{\{m-r,r\}}} & = & \dis\frac{r(r-1)(m-2r+3)}{
m(m-1)(m-2r+1)}\;.
\earr \label{eq.bege-su2-4}
\ee
Using Eqs. (\ref{eq.bege-su2-1})-(\ref{eq.bege-su2-4}) and the expressions in
Table \ref{xuuyuu}, it is possible to derive analytical formulas for the $P$'s, 
$\cq$'s and $\car$'s that define $\lan H^2\ran$, $\hat{\Sigma}_{11}$ and
$\hat{\Sigma}_{22}$. The final formulas (obtained using MATHEMATICA) are, with
$(m,F)$ defining $f_m$,
\be
\barr{l}
P^{\{2\}}(m,F)=-\dis\frac{1}{8}\l[3m(m-2)+4F(F+1)\r]\;,\\ 
P^{\{1^2\}}(m,F)=-\dis\frac{1}{8}\l[m(m+2)-4F(F+1)\r]\;,\\ 
\cq^{\nu=0}(\{2\}:m,F)=\l[P^{\{2\}}(m,F)\r]^2\;,\\ 
\cq^{\nu=0}(\{1^2\}:m,F)=\l[P^{\{1^2\}}(m,F)\r]^2\;,\\ 
\cq^{\nu=1}(\{2\}:m,F)=\dis\frac{(\Omega+1)}{16(\Omega+2)}\;\\
\times \l[2(\Omega-2) P^{\{2\}}(m,F)\l\{3(2\Omega+m)(m-2)+4F(F+1)\r\}
+8\Omega(m-1)(\Omega+2m-4)F(F+1)\r]\;,\\
\cq^{\nu=1}(\{1^2\}:m,F)=\dis\frac{(\Omega-1)P^{\{1^2\}}(m,F)}{8}\;\l[
(2\Omega+m)(m+2)-4F(F+1)\r]\;,\\
\cq^{\nu=2}(\{2\}:m,F)=\dis\frac{(\Omega)}{8(\Omega+2)}\;\l[(3\Omega^2 +
7\Omega +6)[F(F+1)]^2 \r.\\
+\dis\frac{3}{16}\;m(m-2)(2\Omega+m)(2\Omega+m+2)(\Omega-1)(\Omega-2)\;\\
\l. + \dis\frac{F(F+1)}{2} \l\{m(2\Omega+m)(5\Omega+3)(\Omega-2)+ 2\Omega
(\Omega^2-1)(\Omega-6)\r\}\r]\;,\\
\cq^{\nu=2}(\{1^2\}:m,F)=\dis\frac{\Omega(\Omega-3)P^{\{1^2\}}(m,F)}{16}\;\l[
(2\Omega+m)(2\Omega+m-2)-4F(F+1)\r]\;,\\
\car^{\nu=0}(m,F) = P^{\{2\}}(m,F)\,P^{\{1^2\}}(m,F)\;,\\
\car^{\nu=1}(m,F) = \dis\sqrt{\dis\frac{\Omega^2-1}{\Omega^2-4}}\;
\dis\frac{(2-\Omega)P^{\{1^2\}}(m,F)}{8} \l\{4[F(F+1)-3\Omega] +3m(2\Omega+m-2)
\r\}\;.
\earr \label{eq.bege-su2-5}
\ee
Note that Eq. (\ref{eq.bege-su2-5}) is related to the EGUE(2)-$SU(2)$ results
given in \cite{Ko-07}  by the $\Omega \rightarrow -\Omega$ transformation and
they are also  closely  related to the results for spectral variances given in
\cite{Ma-12}.

\begin{figure}[ht]
\centering
    
\includegraphics[width=3in,height=4in]{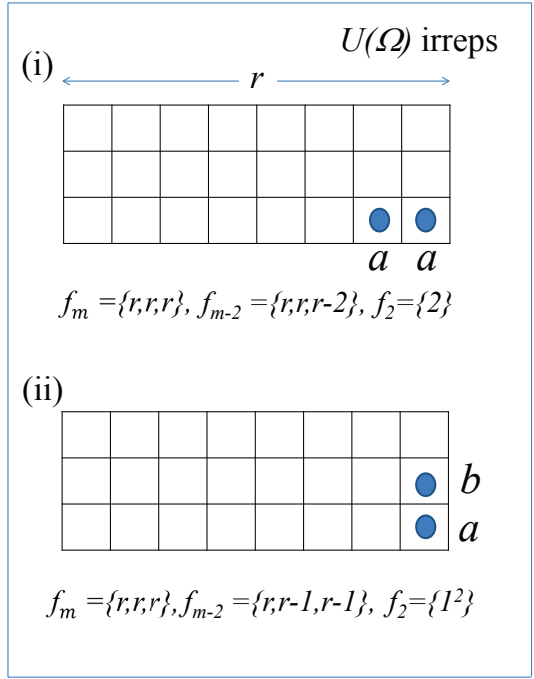}
    
\caption{Young tableaux denoting the three-column $SU(\Omega)$ irreps $f_m = 
\{r,r,r \}$, $m=3r$ appropriate for BEGUE(2)-$SU(2)$. Removal of two boxes
generating $m-2$  particle irreps $f_{m-2}$ are also shown in the figure. For
(i) only the irrep $f_2=\{2\}$ will apply while for (ii) only $\{1^2\}$ will
apply.}

\label{young-su3-bose}
\end{figure}
\begin{figure}
\centering
    
\includegraphics[width=6.25in,height=4.5in]{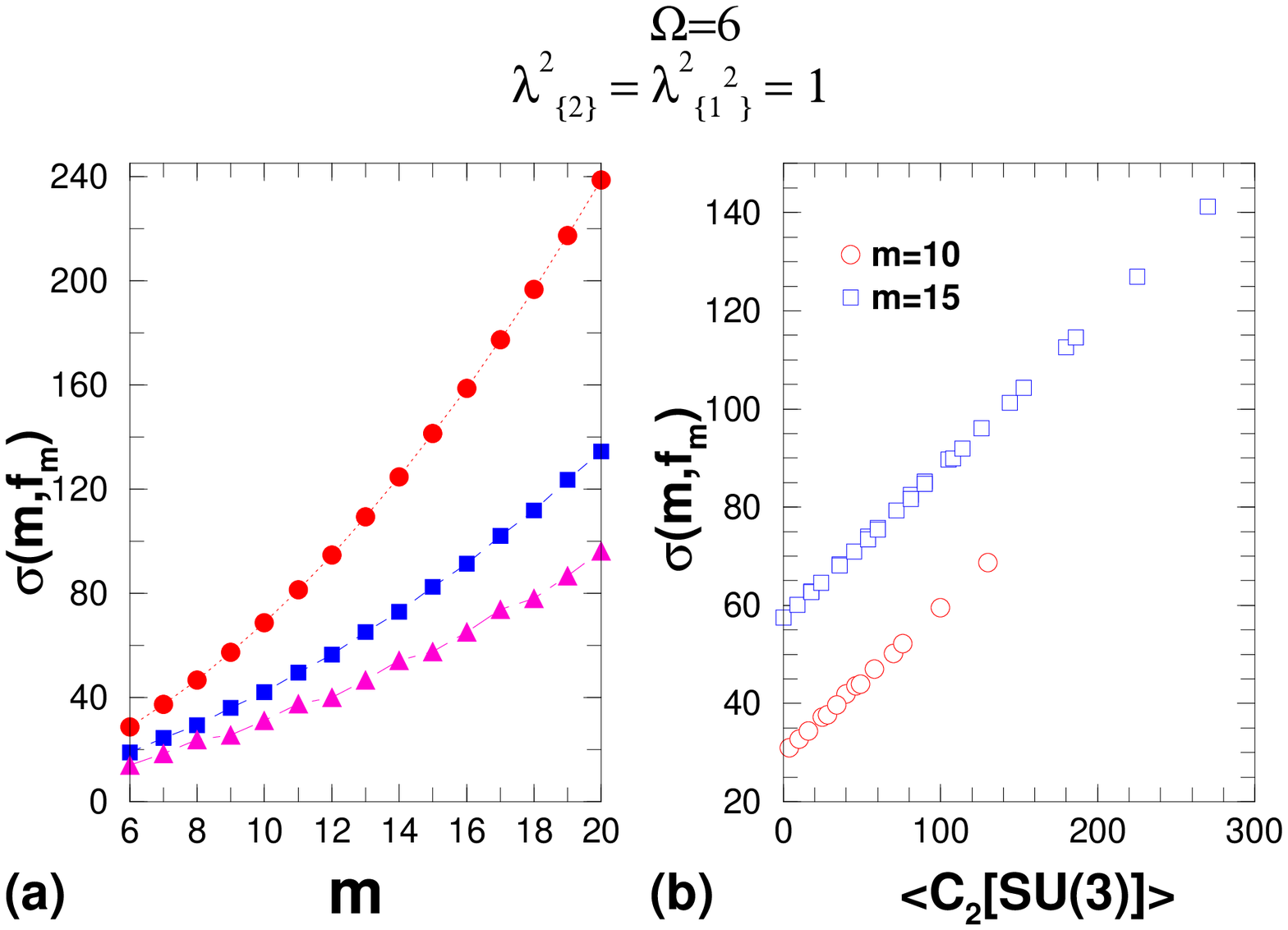}
    
\caption{(a) Variation of spectral widths as a function of $m$ with fixed $f_m$.
Shown are the results for $f_m$ as one-rowed $f_m^{(k)}$ irreps (red circles),
two-rowed $f_m^{(k)}$ irreps (blue squares) and three-rowed $f_m^{(k)}$ irreps
(magenta triangles). (b) Variation of spectral widths as a function of $f_m$
with fixed $m$. Shown are the results for $\Omega=6$ and $m=10,\;15$. Instead of
showing $f_m$, we have used $\lan C_2[SU(3)] \ran^{\tilde{f}_m}$.}

\label{figa}
\end{figure}
   
\section{Embedded Gaussian Unitary Ensemble for spin one bosons: BEGUE(2)-SU(3)
with $r=3$}

Spin one boson systems, discussed in \cite{Ckmp-arx}, possess $U(3\Omega) \supset
U(\Omega) \otimes [SU(3) \supset SO(3)]$ symmetry. Instead of BEGOE(2) or
BEGUE(2) generated by random two-body interactions preserving total spin $S$, it
is also possible to consider interactions preserving the $SU(3)$ symmetry. 
Then, for the GUE version, we have BEGUE(2)-$SU(3)$ that corresponds to $r=3$ 
in Section 2. As $U(3)$ irreps will have, in young tableaux representation,
maximum 3 rows, the $U(\Omega)$ irrep also will have maximum three rows. Given
$m$ bosons in $\Omega$ number of sp levels, the allowed $U(\Omega)$ irreps are
$\{f_1,f_2,f_3,f_4,\ldots,f_\Omega \}$ with $f_1+f_2+f_3=m$, $f_1 \geq f_2 \geq 
f_3 \geq 0$ and $f_i=0$ for $i=4,5,\ldots,\Omega$. Because of the last condition
we use simply  $\{f_1,f_2,f_3\}$. For $f_2=0$ and $f_3=0$, we have totally
symmetric irreps with $\{f_1\}=\{m\}$ and then all the results derived in
Section 3 will apply directly. Similarly, for $f_2 \neq 0$ and $f_3=0$, all
the results of Section 4 will apply. Thus the non-trivial irreps for
BEGUE(2)-$SU(3)$ are the $m$-boson irreps $f_m=\{f_1,f_2,f_3\}$ with 
$f_3 \neq 0$. Given a $f_m$ in general there will be six $f_{m-2}$ and they are
$\{f_1-2,f_2,f_3\}$, $\{f_1,f_2-2,f_3\}$, $\{f_1,f_2,f_3-2\}$, $\{f_1-1,f_2-1,
f_3\}$, $\{f_1-1,f_2,f_3-1\}$, $\{f_1,f_2-1,f_3-1\}$. Therefore, as seen from
Section 2, deriving analytical formulas for $P$'s, $\cq$'s and $\car$'s that
determine $\lan H^2\ran$, $\hat{\Sigma}_{11}$ and $\hat{\Sigma}_{22}$ will be
cumbersome. One situation that is amenable to analytical treatment is for the
irreps $\{r,r,r\}$, $m=3r$. For this class of irreps, the $f_{m-2}$ are simple
as shown in Fig. \ref{young-su3-bose}. For $f_{m-2}=\{r,r,r-2\}$ we need
$\Pi^{\pr}_a$ and $\Pi^{\pr \pr}_a$ and they are given by,
\be
\Pi^{\pr}_a =  \dis\frac{3r}{\Omega+r-3}\;,\;\;\;\Pi^{\pr \pr}_a =  
\dis\frac{6r(r-1)}{(\Omega+r-3)(\Omega+r-4)}\;.
\label{eq.begue-su3-1}
\ee
Similarly, for $f_{m-2}=\{r,r-1,r-1\}$ we need $\tau_{ab}$, $\Pi^{(b)}_a$ and
$\Pi^{(a)}_b$ and they are,
\be
\tau_{ab}=-1\;,\;\;\;\Pi^{(b)}_a = \dis\frac{3r}{2(\Omega+r-3)}\;,\;\;\;
\Pi^{(a)}_b = \dis\frac{2(r+1)}{(\Omega+r-2)} \;.
\label{eq.begue-su3-2}
\ee
In addition, ratio of the $S_\Omega$ dimensions needed are,
\be
\dis\frac{\cn_{r,r,r-2}}{\cn_{r,r,r}} = \dis\frac{2(r-1)}{(3r-1)}\;,\;\;\;\;\;
\dis\frac{\cn_{r,r-1,r-1}}{\cn_{r,r,r}} = \dis\frac{r+1}{(3r-1)}\;.
\label{eq.begue-su3-3}
\ee
With these, carrying out simplification of the formulas given in Table
\ref{xuuyuu} will give the following results,
\be
\barr{l}
P^{\{2\}}(m,\{r,r,r\})=-3r(r-1)\;,\;\;\;P^{\{1^2\}}(m,\{r,r,r\})=-\dis\frac{3}{2}
r(r+1)\;,\\
\cq^{\nu=0}(\{2\}: m,\{r,r,r\})=(3r)^2 (r-1)^2\;,\\
\cq^{\nu=0}(\{1^2\}: m,\{r,r,r\}) =\dis\frac{(3r)^2 (r+1)^2}{4}\;,\\
\cq^{\nu=1}(\{2\}:m,\{r,r,r\}) = \dis\frac{6(\Omega+1)(\Omega-3)r(r-1)^2
(\Omega+r)}{(\Omega+2)}\;,\\
\cq^{\nu=1}(\{1^2\}:m,\{r,r,r\}) = \dis\frac{3(\Omega-1)(\Omega-3)r(r+1)^2
(\Omega+r)}{2(\Omega-2)}\;,\\
\cq^{\nu=2}(\{2\}:m,\{r,r,r\}) =\dis\frac{3 \Omega (\Omega-2)(\Omega-3) r (r-1)
(\Omega+r)(\Omega+r+1)}{4(\Omega+2)}\;,\\
\cq^{\nu=2}(\{1^2\}:m,\{r,r,r\}) =\dis\frac{3 \Omega (\Omega-3)(\Omega-4) r(r+1)
(\Omega+r)(\Omega+r-1)}{8(\Omega-2)}\;,\\
\car^{\nu=0}(m,\{r,r,r\}) = \dis\frac{(3r)^2 (r^2-1)}{2}\;,\\
\car^{\nu=1}(m,\{r,r,r\}) = -\dis\sqrt{\dis\frac{\Omega^2-1}{\Omega^2-4}}\;
3(\Omega-3)r(r^2-1)(\Omega+r)\;.
\earr \label{eq.begue-su3-4}
\ee
Using these equations one can calculate the variances $\lan H^2\ran$ and the
covariances $\hat{\Sigma}_{11}$ and $\hat{\Sigma}_{22}$ for irreps of the type
$\{r,r,r\}$. For example, Eq. (\ref{eq.egsu418}) can be simplified using Eq. (\ref{eq.begue-su3-4}) to give a 
compact formula for spectral variances,
\be
\barr{rcl}
\overline{\lan H^2\ran^{m,\{r,r,r\}}} & = & \lambda^2_{\{2\}}\l[\dis\frac{3}{2}
r(r-1)(\Omega+r-3)(\Omega+r-4)\r] \\ \\
& + & \lambda^2_{\{1^2\}}\l[\dis\frac{3}{4}
r(r+1)(\Omega+r-2)(\Omega+r-3)\r]\;.
\earr \label{eq.begue-su3-5}
\ee 

\begin{figure}[ht]
\centering
    
\includegraphics[width=6.5in,height=5.5in]{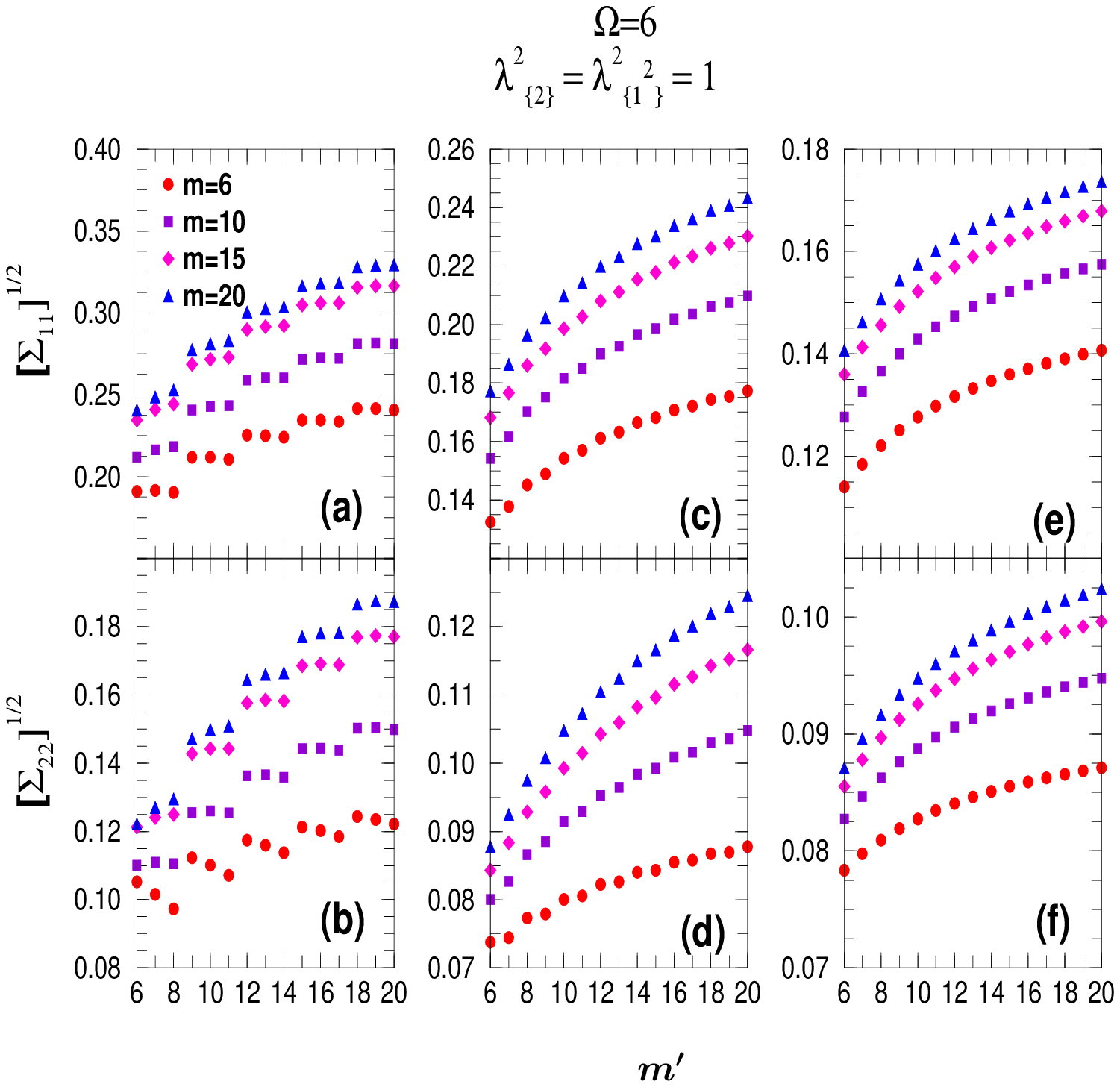}
    
\caption{Self and cross correlations in energy centroids and spectral variances
as a function of $m$ and $m^\pr$ for $\Omega=6$ (with fixed $f_m$ and
$f_{m^\pr}$): (a) $[\Sigma_{11}(m,f_m;m^\pr,f_{m^\pr})]^{1/2}$ with
$\{f_m\}=\{m/3,m/3,m/3\}$ for $m \mod 3 = 0$, 
$\{f_m\}=\{(m+2)/3,(m-1)/3,(m-1)/3\}$ for $m \mod 3 = 1$ and
$\{f_m\}=\{(m+4)/3,(m-2)/3,(m-2)/3\}$ for $m \mod 3 = 2$ and similarly
$f_{m^\pr}$ is defined; (b) $[\Sigma_{22}(m,f_m;m^\pr,f_{m^\pr})]^{1/2}$ with
$\{f_m\}=\{m/3,m/3,m/3\}$ for $m \mod 3 = 0$, 
$\{f_m\}=\{(m+2)/3,(m-1)/3,(m-1)/3\}$ for $m \mod 3 = 1$ and
$\{f_m\}=\{(m+4)/3,(m-2)/3,(m-2)/3\}$ for $m \mod 3 = 2$ and similarly
$f_{m^\pr}$ is defined; (c) $[\Sigma_{11}(m,f_m;m^\pr,f_{m^\pr})]^{1/2}$ with
$\{f_m\}=\{m/2,m/2\}$ for $m \mod 2 = 0$ and  $\{f_m\}=\{(m+1)/2,(m-1)/2\}$ for
$m \mod 2 = 1$ and similarly $f_{m^\pr}$ is defined; (d)
$[\Sigma_{22}(m,f_m;m^\pr,f_{m^\pr})]^{1/2}$ with $\{f_m\}=\{m/2,m/2\}$ for $m
\mod 2 = 0$ and  $\{f_m\}=\{(m+1)/2,(m-1)/2\}$ for $m \mod 2 = 1$ and similarly
$f_{m^\pr}$ is defined; (e) $[\Sigma_{11}(m,f_m;m^\pr,f_{m^\pr})]^{1/2}$ with
$\{f_m\}=\{m\}$ and $\{f_{m^\pr}\}=\{m^\pr\}$; (f)
$[\Sigma_{22}(m,f_m;m^\pr,f_{m^\pr})]^{1/2}$ with $\{f_m\}=\{m\}$ and
$\{f_{m^\pr}\}=\{m^\pr\}$.}

\label{figb}
\end{figure}
\begin{figure}[ht]
\centering
    
\includegraphics[width=6.5in,height=6.75in]{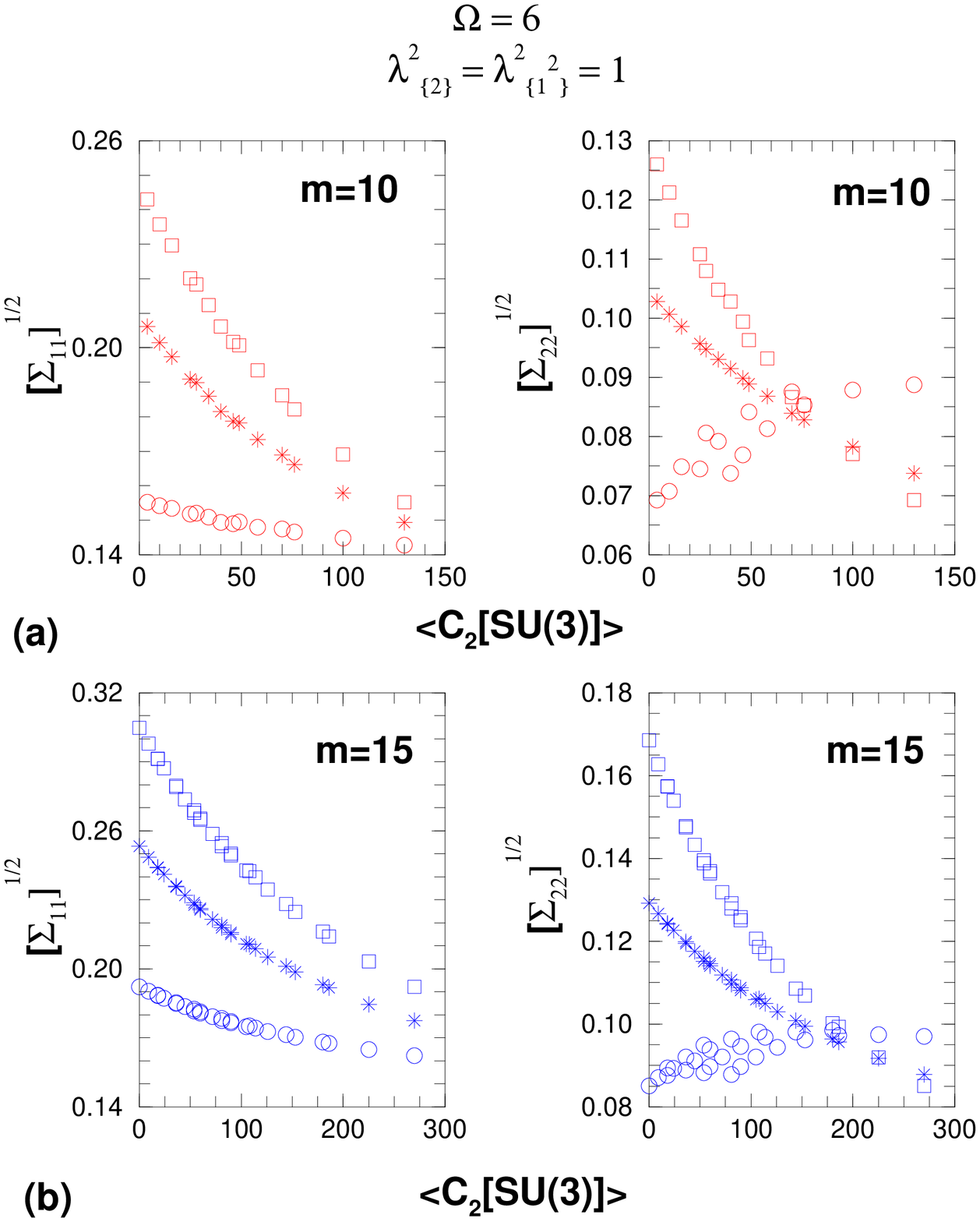}
    
\caption{Self and cross correlations in energy centroids and spectral variances
as a function of $f_m$ and $f_{m^\pr}$ for $\Omega=6$ (with fixed $m=m^\pr$).
Results are shown for (a) $m=m^\pr=10$ with $f_m=\{10\}$ (red circles),
$\{5,5\}$ (red stars) and $\{4,3,3\}$ (red squares) with all one, two and three
rowed $f_{m^\pr}$ irreps; (b) $m=m^\pr=15$ with $f_m=\{15\}$ (blue circles),
$\{8,7\}$ (blue stars) and $\{5,5,5\}$ (blue squares) with all one, two and
three rowed $f_{m^\pr}$ irreps.}

\label{figc}
\end{figure}

It is also possible to derive analytical results for the irreps
$f_m=\{r+1,r,r\}$ and $\{r+2,r,r\}$ just as it was done for $\{4^r,p\}$ irreps
for EGUE(2)-$SU(4)$ ensemble in \cite{Ma-su4}. The results are as follows. For
these irreps, the allowed $f_{m-2}$ irreps and the corresponding $\tau$ and
$\Pi_{\ldots}$ functions are given in  Table \ref{func} and the dimension ratios
in Table \ref{dim}. Using these,  for $f_m=\{r+1,r,r\}$ irreps, $P$, $\cq$ and
$\car$ functions are, 
\be
\barr{l}
P^{\{2\}}(m,\{r+1,r,r\})=- r(3 r-1)
\;,\;\;\;
P^{\{1^2\}}(m,\{r+1,r,r\})= -\dis\frac{r}{2} (5+3 r)
\;,\\
\cq^{\nu=0}(\{2\}: m,\{r+1,r,r\})= r^2 (3 r-1)^2 
\;,\\
\cq^{\nu=0}(\{1^2\}: m,\{r+1,r,r\}) = \dis\frac{r^2}{4} (5+3 r)^2
\;,\\
\cq^{\nu=1}(\{2\}:m,\{r+1,r,r\}) = \dis\frac{r (1+\Omega )}{2+\Omega }
 \l[ 6 r^3 (-3+\Omega )+3 (-3+\Omega ) \Omega  \r. \\ \l.
+2 r^2 (-3+\Omega ) (-2+3 \Omega )
+r \l\{-2+3 (9-2 \Omega ) \Omega \r\} \r]
\;,\\
\cq^{\nu=1}(\{1^2\}:m,\{r+1,r,r\}) = \dis\frac{r (-1+\Omega )}{2 (-2+\Omega )}
 \l[-r (5+3 r)^2 \r. \\ \l.
+\l\{-18+r (-18+r+3 r^2)\r\} \Omega +3 (1+r) (2+r) \Omega ^2 \r]
\;,\\
\cq^{\nu=2}(\{2\}:m,\{r+1,r,r\}) = \dis\frac{r (-3+\Omega ) \Omega  (1+r+\Omega ) }{4 (2+\Omega )}
\l[ 8+3 r^2 (-2+\Omega ) \r. \\ \l.
-(-8+\Omega ) \Omega +r (-2+\Omega ) (1+3 \Omega ) \r]
\;,\\
\cq^{\nu=2}(\{1^2\}:m,\{r+1,r,r\}) = \dis\frac{r (-3+\Omega ) \Omega  (-1+r+\Omega )}{8 (-2+\Omega )}
\l[-16+3 r^2 (-4+\Omega ) \r. \\ \l.
+r (-4+\Omega ) (7+3 \Omega )+\Omega  (-14+5 \Omega )\r]
\;,\\
\car^{\nu=0}(m,\{r+1,r,r\}) = \dis\frac{ r^2}{2} (-1+3 r) (5+3 r)
\;,\\
\car^{\nu=1}(m,\{r+1,r,r\}) = \dis\frac{r}{24} \dis\sqrt{\dis\frac{\Omega ^2-1}{\Omega ^2-4}}
 \l[r^3 (468-151 \Omega )+153 (-3+\Omega ) \Omega \r. \\ \l.
+r^2 \l\{600+(283-151 \Omega ) \Omega \r\}
-5 r \l\{60+13 \Omega  (-7+2 \Omega )\r\}\r]
\;.
\earr \label{ee.begue-su3-5}
\ee
Similarly, for $f_m=\{r+2,r,r\}$ irreps we have,
\be
\barr{l}
P^{\{2\}}(m,\{r+2,r,r\})= -(3 r^2+r+1)
\;,\;\;\;
P^{\{1^2\}}(m,\{r+2,r,r\})= -\dis\frac{r}{2} (7+3 r)
\;,\\
\cq^{\nu=0}(\{2\}: m,\{r+2,r,r\})= (3 r^2+r+1)^2
\;,\\
\cq^{\nu=0}(\{1^2\}: m,\{r+2,r,r\}) = \dis\frac{ r^2}{4} (7+3 r)^2
\;,\\
\cq^{\nu=1}(\{2\}:m,\{r+2,r,r\}) = \dis\frac{(1+\Omega )}{2 (2+\Omega )} \l[-4 (1+r+3 r^2)^2
\r. \\ \l.
+\l\{ 2+r \l\{3+r \l\{51+4 r (-7+3 r)\r\}\r\}\r\} \Omega +(2+11 r+12 r^3) \Omega ^2 \r]
\;,\\
\cq^{\nu=1}(\{1^2\}:m,\{r+2,r,r\}) = \dis\frac{r (-1+\Omega )}{2 (-2+\Omega )}
 \l[-r (7+3 r)^2 \r. \\ \l.
+(5+3 r) (-6+r^2) \Omega +\l\{10+3 r (4+r)\r\} \Omega ^2 \r]
\;,\\
\cq^{\nu=2}(\{2\}:m,\{r+2,r,r\}) = \dis\frac{\Omega}{4 (2+\Omega )}  
\l[ 3 r^4 (-3+\Omega ) (-2+\Omega )+(-1+\Omega ) \Omega  (2+\Omega ) (3+\Omega ) \r. \\ \l.
+2 r^3 (-3+\Omega ) (-2+\Omega ) (4+3 \Omega )
+r^2 (-3+\Omega ) \Omega  \l\{7+3 \Omega  (1+\Omega )\r\} \r. \\ \l.
+r (-3+\Omega ) \l\{22+\Omega  \l\{42+\Omega  (19+\Omega )\r\}\r\} \r]
\;,
\earr
\ee
\be
\barr{l}
\cq^{\nu=2}(\{1^2\}:m,\{r+2,r,r\}) = \dis\frac{r (-3+\Omega ) \Omega  (-1+r+\Omega )}{8 (-2+\Omega )}
\l[ -44+3 r^2 (-4+\Omega ) \r. \\ \l.
+r (-4+\Omega ) (11+3 \Omega )+\Omega  (-12+7 \Omega ) \r]
\;,\\
\car^{\nu=0}(m,\{r+2,r,r\}) = \dis\frac{r}{2} (7+3 r) (1+r+3 r^2)
\;,\\
\car^{\nu=1}(m,\{r+2,r,r\}) = \dis\frac{r}{48} \dis\sqrt{\dis\frac{\Omega ^2-1}{\Omega ^2-4}} 
\l[ 8 (7+3 r) (12+r+37 r^2) \r. \\ \l.
+\l\{-1008+r \l\{1106+(241-289 r) r\r\}\r\} \Omega +\l\{208-r (489+289 r)\r\} \Omega ^2 \r]
\;. \nonumber
\earr \label{ee.begue-su3-6}
\ee

\begin{table}
\caption{Formulas for the functions $\Pi_{--}^{(--)}$'s defined in Eq. \eqref{eq.begue2-2} as required 
for $\{f_m\}$ irreps $\{r+1,r,r\}$ and $\{r+2,r,r\}$. Given also are the values of 
axial distances ($\tau_{--}$'s). Also, $\Pi_a^{(bc)}={r}/{(\Omega+r-3)}$ for both
$\{f_m\}$ examples shown in the Table.}
\begin{tabular}{lll}
\hline \hline
$\{f_m\}$ & $\{f_{m-2}\}$ & \text{Required functions} \\ 
\hline \hline
$\{ r+1,r,r \}$ & $\{r+1,r,r-2\}$ & $\Pi_a^\pr=\dis\frac{8r}{3(\Omega+r-3)}$, $\Pi_a^{\pr\pr}=\dis\frac{5r(r-1)}{(\Omega+r-3)(\Omega+r-4)}$ \\
 & $f(aa)$ & \\ 
\hline
 & $\{r+1,r-1,r-1\}$ & $\tau_{ab}=-1$, $\Pi_a^{(b)}=\dis\frac{4r}{3(\Omega+r-3)}$, $\Pi_b^{(a)}=\dis\frac{3(r+1)}{2(\Omega+r-2)}$ \\
 & $f(ab)$ & \\
\hline
 & $\{r,r,r-1\}$ & $\tau_{ac}=-3$, $\Pi_a^{(c)}=\dis\frac{2r}{\Omega+r-3}$, $\Pi_c^{(a)}=\dis\frac{r+3}{2(\Omega+r)}$ \\
 & $f(ac)$ & \\
\hline
$\{r+2,r,r\}$ & $\{r+2,r,r-2\}$ & $\Pi_a^\pr=\dis\frac{5r}{2(\Omega+r-3)}$, $\Pi_a^{\pr\pr}=\dis\frac{9r(r-1)}{2(\Omega+r-3)(\Omega+r-4)}$ \\
 & $f(aa)$ & \\
\hline
 & $\{r+2,r-1,r-1\}$ & $\tau_{ab}=-1$, $\Pi_a^{(b)}=\dis\frac{5r}{4(\Omega+r-3)}$, $\Pi_b^{(a)}=\dis\frac{4(r+1)}{3(\Omega+r-2)}$ \\ 
 & $f(ab)$ & \\ 
\hline
 & $\{r+1,r,r-1\}$ & $\tau_{ac}=-4$, $\Pi_a^{(c)}=\dis\frac{2r}{\Omega+r-3}$, $\Pi_c^{(a)}=\dis\frac{2(r+4)}{3(\Omega+r+1)}$ \\
 & $f(ac)$ & \\
\hline
 & $\{r,r,r\}$ & $\Pi_c^\pr=\dis\frac{r+4}{2(\Omega+r+1)}$, 
$\Pi_c^{\pr\pr}=\dis\frac{(r+3)(r+4)}{6(\Omega+r)(\Omega+r+1)}$ \\
 & $f(cc)$ & \\
\hline \hline
\end{tabular}
\label{func}
\end{table}

\begin{table}
\caption{Dimension ratios with respect to the $S_m$ group for the examples in Table \ref{func}.}
\begin{tabular}{lll}
\hline\hline 
$\{f_m\}$ & $\{f_{m-2}\}$ & $\dis\frac{\cn_{\{f_{m-2}\}}}{\cn_{\{f_m\}}}$ \\
\hline\hline
$\{ r+1,r,r \}$ & $\{r+1,r,r-2\}$ & $\dis\frac{5(r-1)}{3(3r+1)}$ \\
 & $\{r+1,r-1,r-1\}$ & $\dis\frac{2(r+1)}{3(3r+1)}$ \\
 & $\{r,r,r-1\}$ & $\dis\frac{(r+3)}{3(3r+1)}$ \\
\hline
$\{r+2,r,r\}$ & $\{r+2,r,r-2\}$ & $\dis\frac{9r(r-1)}{2(3r+2)(3r+1)}$ \\
 & $\{r+2,r-1,r-1\}$ & $\dis\frac{5r(r+1)}{3(3r+2)(3r+1)}$ \\ 
 & $\{r+1,r,r-1\}$ & $\dis\frac{4r(r+4)}{3(3r+2)(3r+1)}$ \\
 & $\{r,r,r\}$ & $\dis\frac{(r+3)(r+4)}{6(3r+2)(3r+1)}$ \\
\hline\hline
\end{tabular}
\label{dim}
\end{table}

In addition to analytical results, as stated in Section 2, one can use the
tables in \cite{Ma-su4} (also to a large extent Table \ref{xuuyuu}) and obtain
numerical results for the variation of spectral variances with the eigenvalues
of the quadratic Casimir invariant of $U(\Omega)$ or equivalently $C_2[SU(3)]$,
for various $(\Omega,m)$ values and also for both self and cross correlations in
energy centroids and spectral variances. Note that For a $\Omega=6$ system with
$\lambda_{\{2\}}^2 = \lambda_{\{1^2\}} = 1$  calculations are carried out for
various choices of $m$ and $f_m$ and the results are shown in \ref{figa},
\ref{figb} and \ref{figc}. Let us mention that $C_2[SU(3)]$ for a irrep
$\{f_1,f_2,f_3\}$ is given by the formula,
\be
\lan C_2[SU(3)]\ran^{\{f_1,f_2,f_3\}} = \lambda^2 + \mu^2 + \lambda \mu + 
3(\lambda+\mu)\;;\;\;\;\lambda=f_1-f_2,\;\mu=f_2-f_3 \;.
\label{sue-casim}
\ee
It is seen from Fig. \ref{figa}a that the spectral widths will be largest for
one rowed irreps and smallest for three row irreps for a fixed $m$. Also, widths
as expected increase with $m$. Similarly, Fig. \ref{figa}b shows that for a
fixed $m$, widths increase as the eigenvalue of $C_2[SU(3)]$ increases and this is
consistent with the observation in Fig. \ref{figa}a as the eigenvalue of
$C_2[SU(3)]$ is largest for totally symmetric irrep. Results in Fig. \ref{figb}
show that: (i) the centroid and variance fluctuations increase with $m^\pr$ for
fixed $m$ and vice-versa; (ii) they are larger for three rowed irreps compared to 
those for one rowed irreps; (iii) centroid fluctuations are much larger than variance
fluctuations as seen before also for EGUE(2)-$\cs$ and EGUE(2)-$SU(4)$
ensembles. Similar trends are also seen for $m=m^\pr$ but varying $f_{m^\pr}$
with fixed $f_m$ and these results are shown in Fig. \ref{figc}. A different
trend is seen for the covariances in spectral variances for the totally symmetric irrep
$f_m=\{m\}$ and $f_{m^\pr}$ varying. More importantly,
the centroid and variance fluctuations are smallest for the ground state i.e., the
most symmetric irrep for bosons. It is seen from Figs. \ref{figb} and \ref{figc} 
that the covariances in energy centroids are $\sim 15-25$\% and the covariances 
in spectral variances are $\sim 8-15$\%.
 
\section{Conclusions}

In this paper, given first is a general formulation for deriving lower order 
moments of the one- and two-point correlation functions in eigenvalues that is
valid for any embedded random matrix for fermions as well as for bosons with $U(\Omega)
\otimes SU(r)$ embedding and with two-body interactions preserving $SU(r)$
symmetry. Results of the present paper unify all the results known before for EGUE(2)'s and 
BEGUE(2)'s. Presented are new results for boson systems with $SU(r)$ symmetry, $r=2,3$. 
These results should be useful in future studies of two species boson systems and spin 
one boson systems. In future, it will be useful to derive analytical forms for $SU(\Omega)$
Racah coefficients \cite{Ko-07,Ma-su4} or develop tractable methods for their numerical evaluation
to establish Gaussian form of the eigenvalue densities generated by embedded ensembles 
with $SU(r)$ symmetry both for boson and fermion systems.

\acknowledgments

Thanks are due to N.D. Chavda for some useful discussions. 
M. V.  gratefully acknowledges financial support from the US 
National Science Foundation grant PHY-0855337.

\ed